\UseRawInputEncoding
\documentclass[prd,showpacs,floatfix,amsmath,amssymb,floatfix,nofootinbib]{revtex4}
\usepackage{graphicx,color,dcolumn,booktabs,bm}
\usepackage{longtable,lscape}
\usepackage{txfonts}
\usepackage{overpic}
\usepackage{amssymb}
\usepackage{indentfirst}
\usepackage{mathrsfs}
\usepackage{amsfonts}
\usepackage{amsmath}
\usepackage{array}
\usepackage{verbatim}
\usepackage{epsfig}
\usepackage{graphicx,color}
\usepackage{relsize}
\usepackage{lineno}
\usepackage{multirow}

\RequirePackage{xspace}

\begin{document}

\title{\boldmath Production of the bottomonium-like $Z_b$ states in $e
$-$h$ and ultraperipheral $h$-$h$ collisions}
\author{Xiao-Yun Wang$^{1}$}
\thanks{xywang@lut.edu.cn (Corresponding author)}
\author{Wei Kou$^{2,3}$}
\author{Qing-Yong Lin$^{4}$}
\thanks{qylin@jmu.edu.cn}
\author{Ya-Ping Xie$^{2,3}$}
\thanks{xieyaping@impcas.ac.cn}
\author{Xurong Chen$^{2,3,5}$}
\thanks{xchen@impcas.ac.cn}
\affiliation{$^1$Department of physics, Lanzhou University of Technology, Lanzhou 730050,
China\\
$^2$Institute of Modern Physics, Chinese Academy of Sciences, Lanzhou
730000, China\\
$^3$University of Chinese Academy of Sciences, Beijing 100049, China\\
$^4$Department of Physics, Jimei University, Xiamen 361021, China\\
$^5$Guangdong Provincial Key Laboratory of Nuclear Science, Institute of Quantum Matter, South China Normal University, Guangzhou 510006, China}
\author{Alexey Guskov$^{6}$}
\thanks{avg@jinr.ru}
\affiliation{$^6$Joint Institute for Nuclear Research, Dubna 141980, Russia}

\begin{abstract}
The photoproduction of bottomonium-like states $Z_{b}(10610)$ and $%
Z_{b}(10650)$ via $\gamma p$ scattering is studied within an effective
Lagrangian approach and the vector-meson-dominance model. The Regge model is employed to calculate
the photoproduction of $Z_{b}$ states via $t$-channel with $\pi$ exchange.
The numerical results show that the values of the total cross-sections of $Z_{b}(10610)$ and $Z_{b}(10650)$ can reach 0.09 nb and 0.02 nb, respectively, near the center of mass energy of 22 GeV. The experimental measurements and studies on the photoproduction of $Z_{b}$ states near energy region around $W\simeq 22$ GeV is suggested. Moreover, with the help of eSTARlight and STARlight programs, one obtains the cross-sections and event numbers of $Z_{b}(10610)$ production in electron-ion collision (EIC) and Ultraperipheral collisions (UPCs). The results show that a considerable number of events from $Z_{b}(10610)$ can be produced on the relevant experiments of EICs and UPCs. Also, one calculates the rates and kinematic distributions for $\gamma p\rightarrow Z_{b}n$ in $ep$ and $pA$ collisions via EICs and UPCs, and the relevant results will provide an important reference for the  RHIC, LHC, EIC-US, LHeC, and FCC  experiments to search for the bottomonium-like $Z_{b}$ states.
\end{abstract}

\maketitle

\section{Introduction}

In recent decades, with the continuous progress of high energy physics
experiments, more and more exotic hadron states have been discovered \cite%
{Tanabashi:2018oca,Liu:2019zoy,Guo:2019twa}. The study of the production and
properties of exotic hadron states is not only conducive to the improvement
and development of hadron spectrum and hadron classification, but also of
great significance for an in-depth understanding of non-perturbative quantum
chromodynamics (QCD). The candidate particles of the exotic states that have
been discovered are mostly concentrated in the charm energy region, and the
discovered exotic states in the bottom quark energy region are still very
limited \cite{Tanabashi:2018oca,Liu:2019zoy,Guo:2019twa}. In 2011, two
bottomonium-like states $Z_{b}(10610)$ and $Z_{b}(10650)$, were observed by
the Belle Collaboration \cite{belle1}, and a series of subsequent
experiments also discovered these two states from different decay channels
\cite{belle1,belle2,belle3,belle4}. Since the quantum numbers and decay
properties of $Z_{b}(10610)$ and $Z_{b}(10650)$ are very similar \cite%
{Tanabashi:2018oca}, for convenience, $Z_{b}(10610)$ and $Z_{b}(10650)$ will
be abbreviated as $Z_{b}$ later. These two states are considered to be
different from the traditional hadron states and are likely to contain at
least four quarks\cite{Liu:2019zoy,Guo:2019twa}.

Observations of the $Z_b$ has inspired extensive studies on the underlying properties, where tetraquark state \cite{Ali:2011ug,Esposito:2014rxa,Maiani:2017kyi,Wang:2013zra},
hadronic molecule interpretation \cite%
{Bondar:2011ev,Zhang:2011jja,Sun:2011uh,Ke:2012gm,Dias:2014pva,Wang:2018jlv}
are performed. More discussions can be found in Refs. \cite%
{Guo:2017jvc,Liu:2019zoy}. In Ref. \cite{Guo:2019twa}, the authors pointed
out that, since these two states are discovered through the decay reaction
of the bottomonium, the contribution of the triangular singularities during
the reaction cannot be neglected, which means that one cannot yet determine
whether these two states are genuine particles. At present, investigating $Z_b$ is still an interesting research topic.

Besides the analysis of the mass spectrum and the decay behavior, studying
the production of $Z_b$ in more different mechanisms is very helpful to
obtain definite evidence for their nature as genuine states. As well
known, the meson photoproduction process was proposed to be an effective way
to search for exotic states \cite%
{Liu:2008qx,He:2009yda,Galata:2011bi,Lin2013,Wang:2015lwa,Wang:2019krd,Wang:2019zaw,Xie:2020wfe,Albaladejo:2020tzt}. We
take notice of the $Z_b \to \Upsilon(nS)\pi^+$ decay modes, which indicates
that there exists a strong coupling between $Z_b$ and $\Upsilon(nS)\pi^+$.
Since $\Upsilon(nS)$ is a vector meson, we suppose that we can carry out
the production of the $Z_b$ states through the meson photoproduction. In the
current work, the photoproduction of $Z_{b}$ will be studied within the
framework of the effective Lagrangian approach and the vector-meson-dominance
(VMD) model \cite{Bauer:1975bv,Bauer:1975bw,Bauer:1977iq}. The calculations
will provide crucial information on the suitable process and the best energy
window of searching for the $Z_b$ states on related photoproduction
experiments.

In hadron-hadron collisions, when the impact parameter between the two nuclei is larger than the
sum of radii of two nuclei, the direct strong interaction between the nuclei is suppressed since the strong interaction is short
range. However, the
electronic-magnetic interaction can not be neglected since it is long-range interaction. This collision
is denoted as ultraperipheral collisions (UPCs) \cite{Bertulani:2005ru,Baltz:2007kq}.
In UPCs, the photon is almost a real photon when the mass number of an atomic nucleus is larger
than 16. Hence, UPCs is a good platform to study photoproduction with small photon virtuality.

Electron-ions collider (EIC) is an important platform to investigate nucleon structure in the future. In EICs, the electron scatters off a nucleon or
nuclei via a virtual photon. Then, vector mesons and exotic states can be produced. Thus, the photoproduction of exotic states can be studied in EICs in the future.
There are a couple of proposed EICs plans
in the world, for example,
EicC, EIC-US, LHeC and FCC are proposed \cite{Chen:2018wyz,Montag:2017,AbelleiraFernandez:2012cc,Bordry:2018gri}.
In EICs, the photon emitted from the electron beam
has large virtuality. This is different from the photon in UPCs. Hence, EICs can be applied to
investigate the photoproduction in a large $Q^2$ region.

STARlight and eSTARlight are two important Monte-Carlo packages to simulate the photoproduction of vector mesons and exotic states
in UPCs and EICs \cite{Klein:2016yzr,Lomnitz:2018juf,Klein:2019avl}. The cross-sections of vector mesons and exotic states produced in photon-proton scattering is
 needed in the simulation process. The information of four-momentum of final states are produced in the simulation processes.
 At the same, the total cross sections of vector mesons or exotic states in UPCs and EICs are performed in STARlight and eSTARlight. In this work, the total cross sections of $Z_b$ at UPCs
 and proposed EICs will be performed adopting STARlight and eSTARlight packages. The rapidity and transverse momentum distributions  of $Z_b$ will be presented. These distributions will be useful for the detector systems in future experiments.

This paper is organized as follows. After the introduction, one presents the
formalism for the production of $Z_{b}$ in Section II. The numerical results
of the $Z_{b}$ production follow in Section III. Finally, the paper ends
with a summary.

\section{Formalism}

\subsection{$Z_{b}$ photoproduction in $\protect\gamma p\rightarrow Z_{b}n$
reaction}

\begin{figure}[tbph]
\begin{center}
\includegraphics[scale=0.6]{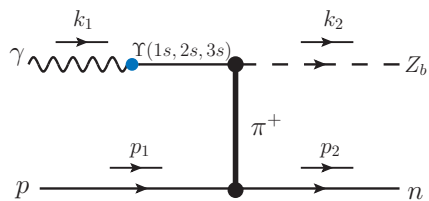}
\end{center}
\caption{Feynman diagrams for the reaction $\protect\gamma p\rightarrow
Z_{b}n$.}
\label{Fig: Feynman1}
\end{figure}

In this work, the production of the hidden-bottom $Z_{b}(10610)$ and $%
Z_{b}(10650)$ states via $\gamma p\rightarrow Z_{b}n$ reaction will be studied with an effective Lagrangian approach. In the PDG
book \cite{Tanabashi:2018oca}, one finds that the $Z_{b}(10610)$ or $%
Z_{b}(10650)$ can decay to a bottomonium plus $\pi $ meson with a branching
ratio of a few percent. Since $Z_{b}$ states are not directly coupled to
photons, the VMD model can be used to calculate the
photoproduction of $Z_{b}$ states through the $t$ channel with $\pi $ exchange.
The Feynman diagram of the $\gamma p\rightarrow Z_{b}n$ reaction via $t$
channel $\pi $ exchange is depicted in Fig. \ref{Fig: Feynman1}. It is noted
from Fig. \ref{Fig: Feynman1} that we only consider the coupling of $%
\Upsilon (1s,2s,3s)$ and photons. Also, although $Z_{b}$ can also
decay to $h_{b}(1p,2p)\pi $, since the parity of the $h_{b}$ state is
opposite to that of the photon, the direct coupling of $%
h_{b}$ to photon can be neglected.

\subsubsection{Lagrangians for the $Z_{b}$ production}

In PDG \cite{Tanabashi:2018oca}, the spin-parity quantum numbers of $%
Z_{b}(10610)$ and $Z_{b}(10650)$ are both $1^{+}$, thus the Lagrangian
densities for the vertices of $Z_{b}\Upsilon \pi $ and $\pi NN$ are written as
\cite{Liu:2008qx,Wang:2019dsi},%
\begin{eqnarray}
\mathcal{L}_{Z_{b}\Upsilon \pi } &=&\frac{g_{Z_{b}\Upsilon \pi }}{M_{Z_{b}}}%
(\partial ^{\mu }\Upsilon ^{\nu }\partial _{\mu }\pi Z_{b\nu }-\partial
^{\mu }\Upsilon ^{\nu }\partial _{\nu }\pi Z_{b\mu }), \\
\mathcal{L}_{\pi NN} &=&-ig_{_{\pi NN}}\bar{N}\gamma _{5}\vec{\tau}\cdot
\vec{\pi}N,
\end{eqnarray}%
where $Z_{b}$, $\Upsilon $, $\pi $ and $N$ denote the fields of $%
Z_{b}(10610)/Z_{b}(10650)$, $\Upsilon $, pion and nucleon meson,
respectively. Here, the $g_{_{\pi NN}}^{2}/4\pi =12.96$ is adopted \cite%
{Lin:1999ve}.

The coupling constant $g_{Z_{b}\Upsilon \pi }$ can be derived
from the corresponding decay width%
\begin{eqnarray}
\Gamma _{Z_{b}\rightarrow \Upsilon \pi } &=&\left( \frac{g_{Z_{b}\Upsilon
\pi }}{M_{Z_{b}}}\right) ^{2}\frac{|\vec{p}_{\pi }^{~\mathrm{c.m.}}|}{24\pi
M_{Z_{b}}^{2}}  \notag \\
&&\times \left[ \frac{(M_{Z_{b}}^{2}-m_{\Upsilon }^{2}-m_{\pi }^{2})^{2}}{2}%
+m_{\Upsilon }^{2}E_{\pi }^{2}\right] ,
\end{eqnarray}%
with%
\begin{eqnarray}
|\vec{p}_{\pi }^{~\mathrm{c.m.}}| &=&\frac{\lambda
^{1/2}(M_{Z_{b}}^{2},m_{\Upsilon }^{2},m_{\pi }^{2})}{2M_{Z_{b}}}, \\
E_{\pi } &=&\sqrt{|\vec{p}_{\pi }^{~\mathrm{c.m.}}|^{2}+m_{\pi }^{2}},
\end{eqnarray}%
where $\lambda $ is the K\"{a}llen function with $\lambda (x,y,z)\equiv
\sqrt{(x-y-z)^{2}-4yz}$, and $M_{Z_{b}}$, $m_{\Upsilon }$, and $m_{\pi }$
are the masses of $Z_{b}$, $\Upsilon $, and pion meson, respectively. The
partial decay widths and coupling constants for $Z_{b}\rightarrow \Upsilon
\pi $ are listed in Table \ref{tab1}.

\renewcommand\tabcolsep{0.26cm} \renewcommand{\arraystretch}{2}
\begin{table}[tbph]
\caption{The values of coupling constants $g_{Z_{b}\Upsilon%
\protect\pi }$ by taking the corresponding decay width of $\Gamma
_{Z_{b}\rightarrow \Upsilon \protect\pi }$ in PDG book
\protect\cite{Tanabashi:2018oca}. Here the unit of width is MeV.}
\label{tab1}%
\begin{tabular}{ccccccc}
\hline\hline
states & $\Gamma _{Z_{b}\rightarrow \Upsilon (1S)\pi }$ & $g_{Z_{b}\Upsilon
(1S)\pi }$ & $\Gamma _{Z_{b}\rightarrow \Upsilon (2S)\pi }$ & $%
g_{Z_{b}\Upsilon (2S)\pi }$ & $\Gamma _{Z_{b}\rightarrow \Upsilon (3S)\pi }$
& $g_{Z_{b}\Upsilon (3S)\pi }$ \\ \hline
$Z_{b}(10610)$ & 0.099 & 0.487 & 0.662 & 3.299 & 0.386 & 9.292 \\
$Z_{b}(10650)$ & 0.019 & 0.206 & 0.161 & 1.468 & 0.184 & 4.916 \\
\hline\hline
\end{tabular}%
\end{table}

The coupling of $Z_{b}$ to the photon can be derived under the VMD mechanism \cite{Bauer:1977iq,Bauer:1975bv,Bauer:1975bw}%
.\ In the VMD mechanism, a real photon can fluctuate
into a virtual vector meson, which subsequently scatters off the target
proton.\ \

The Lagrangian depicting the coupling of the meson $\Upsilon $ with a photon
reads as \cite{Wang:2019krd,Wang:2019zaw}%
\begin{equation}
\mathcal{L}_{\Upsilon \gamma }=-\frac{em_{\Upsilon}^{2}}{f_{\Upsilon}}\Upsilon _{\mu }A^{\mu },
\end{equation}%
where $f_{\Upsilon}$ is the $\Upsilon $ decay constant. Thus one gets the expression for the $\Upsilon \rightarrow
e^{+}e^{-}$ decay width,%
\begin{equation}
\Gamma _{\Upsilon \rightarrow e^{+}e^{-}}=\left( \frac{e}{f_{\Upsilon }}%
\right) ^{2}\frac{8\alpha \left\vert \vec{p}_{e}^{~\mathrm{c.m.}}\right\vert
^{3}}{3m_{\Upsilon}^{2}},
\end{equation}%
where $\vec{p}_{e}^{~\mathrm{c.m.}}$ denotes the three-momentum of an
electron in the rest frame of the $\Upsilon $ meson. $\alpha =e^{2}/4\pi
=1/137$ is the electromagnetic fine structure constant. With the partial
decay width of $\Upsilon (1s,2s,3s)\rightarrow e^{+}e^{-}$ \cite%
{Tanabashi:2018oca}, one gets $e/f_{\Upsilon (1s)}\simeq 0.008$, $%
e/f_{\Upsilon (2s)}\simeq 0.005$ and $e/f_{\Upsilon (3s)}\simeq 0.004$.

\subsubsection{Reggeized $t$ channel}

Since the energy corresponding to the $\gamma p\rightarrow Z_{b}n$ reaction
is above 10 GeV, the Reggeized treatment will be applied to the $t$ channel
process. Usually, one just needs to replace the Feynman propagator with the
Regge propagator as%
\begin{equation}
\frac{1}{t-m_{\pi }^{2}}\rightarrow (\frac{s}{s_{scale}})^{\alpha _{\pi }(t)}%
\frac{\pi \alpha _{\pi }^{\prime }}{\Gamma \lbrack 1+\alpha _{\pi }(t)]\sin
[\pi \alpha _{\pi }(t)]},
\end{equation}%
where the scale factor $s_{scale}$ is fixed at 1 GeV. In addition, the Regge
trajectories of $\alpha _{\pi }(t)$ is written as \cite{Wang:2019dsi},%
\begin{equation}
\alpha _{\pi }(t)=0.7(t-m_{\pi }^{2}).\quad \ \
\end{equation}%
It can be seen that no free parameters have been added after introducing the Regge model.

\subsubsection{Amplitude}

Based on the Lagrangians above, the scattering amplitude for the reaction $%
\gamma p\rightarrow Z_{b}n$ can be constructed as%
\begin{equation}
-i\mathcal{M}_{\gamma p\rightarrow Z_{b}n}=\epsilon _{Z_{b}}^{\mu }(k_{2})%
\bar{u}(p_{2})\mathcal{A}_{\mu \nu }u(p_{1})\epsilon _{\gamma }^{\nu
}(k_{1}),
\end{equation}%
where $u$ is the Dirac spinor of nucleon, and $\epsilon _{Z_{b}}$ and $%
\epsilon _{\gamma }$ are the polarization vectors of $Z_{b}$ meson and
photon, respectively.

The reduced amplitude $\mathcal{A}_{\mu \nu }$ for the $t$ channel $Z_{b}$
photoproduction reads
\begin{eqnarray}
\mathcal{A}_{\mu \nu } &=&-i(\sqrt{2}g_{\pi NN}\frac{g_{Z_{b}\Upsilon \pi }}{%
M_{Z_{b}}}\frac{e}{f_{\Upsilon }})\gamma _{5}[k_{1}\cdot (k_{2}-k_{1})g_{\mu
\nu }-k_{1\mu }(k_{2}-k_{1})_{\nu }]  \notag \\
&&\times \frac{1}{q^{2}-m_{\pi }^{2}}\mathcal{F}_{\pi NN}(q^{2})\mathcal{F}%
_{Z_{b}\Upsilon \pi }(q^{2}),  \label{AmpT1}
\end{eqnarray}

For the $t$-channel meson exchanges \cite%
{Liu:2008qx,He:2009yda,Wang:2015lwa,Wang:2019dsi}, the general form factor $%
\mathcal{F}_{t}(q_{t}^{2})$ consisting of $\mathcal{F}_{Z_{b}\Upsilon \pi
}=(m_{\Upsilon }^{2}-m_{\pi }^{2})/(m_{\Upsilon }^{2}-q_{\pi }^{2})$ and $%
\mathcal{F}_{\pi NN}=(\Lambda _{t}^{2}-m_{\pi }^{2})/(\Lambda
_{t}^{2}-q_{\pi }^{2})$ are taken into account. Here, $q_{\pi }$ and $m_{\pi
}$ are 4-momentum and mass of the $\pi $ meson, respectively. The cutoff $\Lambda _{t}$ will be taken as 0.7 GeV, which is the same as that in Ref. \cite{Liu:2008qx,Galata:2011bi,Lin2013,Wang:2015lwa}.

With the preparation in the previous sections, the differential
cross section in the center of mass (c.m.) frame is written as
\begin{equation}
\frac{d\sigma }{d\cos \theta }=\frac{1}{32\pi s}\frac{\left\vert \vec{k}%
_{2}^{{~\mathrm{c.m.}}}\right\vert }{\left\vert \vec{k}_{1}^{{~\mathrm{c.m.}}%
}\right\vert }\left( \frac{1}{4}\sum\limits_{\lambda }\left\vert \mathcal{M}%
\right\vert ^{2}\right) ,
\end{equation}%
Here, $s=(k_{1}+p_{1})^{2}$, and $\theta $ denotes the angle of the outgoing
$Z_{b}$ meson relative to $\gamma $ beam direction in the c.m. frame. $\vec{k%
}_{1}^{{~\mathrm{c.m.}}}$ and $\vec{k}_{2}^{{~\mathrm{c.m.}}}$ are the
three-momenta of the initial photon beam and final $Z_{b}$ meson,
respectively.

\subsection{$Z_{b}$ production in EIC and UPCs}

In the electron-proton scattering, the cross section of $Z_{b}$ is given by
\cite{Lomnitz:2018juf,Klein:2019avl}
\begin{equation}
\sigma (ep\rightarrow eZ_{b}n)=\int dkdQ^{2}\frac{dN^{2}(k,Q^{2})}{dkdQ^{2}}%
\sigma _{\gamma ^{\ast }p\rightarrow Z_{b}n}(W,Q^{2}),
\end{equation}%
where $k$ is the momentum of the photon emitted from electron in target rest
frame, $W$ is the c.m. energy of the photon and proton
system, and $Q^{2}$ is the virtuality of the photon. The photon flux reads
as \cite{Budnev:1974de}
\begin{equation}
\frac{d^{2}N(k,Q^{2})}{dkdQ^{2}}=\frac{\alpha }{\pi kQ^{2}}\Big[1-\frac{k}{%
E_{e}}+\frac{k^{2}}{2E_{e}^{2}}-\Big(1-\frac{k}{E_{e}}\Big)\Big|\frac{%
Q_{min}^{2}}{Q^{2}}\Big|\Big].
\end{equation}%
The $Q^{2}$ dependence of $\sigma _{\gamma ^{\ast }p\rightarrow
Z_{b}n}(W,Q^{2})$ is factorized as
\begin{equation}
\sigma _{\gamma ^{\ast }p\rightarrow Z_{b}n}(W,Q^{2})=\sigma _{\gamma
p\rightarrow Z_{b}n}(W,Q^{2}=0)\bigg(\frac{M_{V}^{2}}{M_{V}^{2}+Q^{2}}\bigg)%
^{\eta },
\end{equation}%
where $M_V$ is mass of the vector meson.
Since there is no parameter for $Z_{b}$ state, we apply the same $\eta $
from $J/\psi $ as Ref. \cite{Lomnitz:2018juf}. This assumption has very
little impact on the project for $Z_{b}$ because we consider the the $0<Q^{2}<1%
\mathrm{GeV}^{2}$, which is quasi-real events\cite{Gryniuk:2020mlh}.

The cross-section of an exotic charged particle in UPCs is computed in
integrating the photon flux and photon-proton cross-section. The photon flux
represents the number as a function momentum of the photon emitted from a nucleus. In $p\mbox{-}A$ UPCs, the cross section of the $pA\rightarrow
nAZ_{b} $ reads \cite{Klein:2016yzr}
\begin{equation}
\sigma (pA\rightarrow AZ_{b}n)=\int dk\frac{dN_{\gamma }(k)}{dk}\sigma
_{\gamma p\rightarrow Z_{b}n}(W).  \label{eqcs}
\end{equation}%
where $k$ is the momentum of the real photon emitted from nucleus, $W$ is the
c.m. energy of the photon and proton system. The photon flux of the photon
emitted from nucleus is given as \cite{Klein:1999qj}
\begin{equation}
\frac{dN_{\gamma }(k)}{dk}=\frac{2Z^{2}\alpha }{\pi k}\big(XK_{0}(X)K_{1}(X)-%
\frac{X^{2}}{2}[K_{1}^{2}(X)-K_{0}^{2}(X)]\big).  \label{phflux}
\end{equation}%
where $X=b_{min}k/\gamma _{L}$, $b_{min}=R_{A}+R_{p}$ is sum of radii of
proton and nucleus. $\gamma _{L} = \sqrt{s}/2m_p$ is the Lorentz boost factor. $K_{0}(x)$
and $K_{1}(x)$ are the modified Bessel functions. $Z$ denotes the charge number of the
nucleus.

Adopting the cross-sections of $Z_b$ in photon-proton
interaction, we can obtain the $Z_b$ cross-sections in $e$-$p$
scattering in EIC and $p\mbox{-}A$ UPCs. With the help of
eSTARlight and STARlight packages, we can simulate the $Z_{b}$ production processes
and get the four-momentum of final states. Then, we will further obtain the $Z_{b}$ in rapidity distributions and transverse momentum distributions, where the rapidity is defined as $\mathrm{y} = \frac{1}{2}\ln[(E+p_z)/(E-p_z)]$.

\section{Numerical results}

\subsection{Photoproduction of $Z_{b}$}

\begin{figure}[h]
\begin{center}
\includegraphics[scale=0.41]{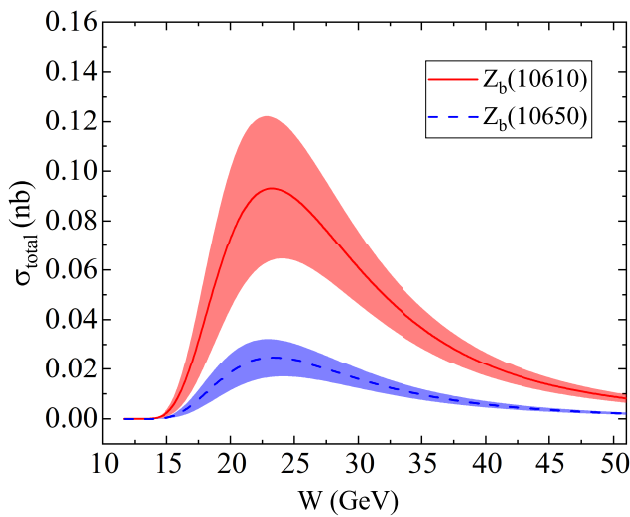}
\end{center}
\caption{The total cross section for the $\protect\gamma p\rightarrow Z_{b}n$
reaction via pionic Regge trajectory exchange. The red solid, and blue
dashed lines are for the $Z_{b}(10610)$, and the $Z_{b}(10650)$,
respectively. Here, the value of cutoff $\Lambda _{t}$ is taken as $0.7\pm
0.2$ GeV. The bands stand for the error bar of the cutoff $\Lambda _{t}.$}
\label{tcs}
\end{figure}

In Fig.~\ref{tcs} we present the total cross sections for the reaction $%
\gamma p\rightarrow Z_{b}n$ from threshold to 50 GeV of the c.m.
energy. One finds that the total cross-section of the $\gamma p\rightarrow
Z_{b}(10610)n$ scattering process reaches a maximum at the center of mass
energy $W\simeq 22$ GeV, which is about 0.09 nb. Also, one notice that when the cutoff parameter
is changed from 0.5 to 0.9 GeV, the total cross section of the $\gamma p\rightarrow
Z_{b}(10610)n$ at $W\simeq 22$ GeV is approximately between 0.06 and 0.12 nb.

\subsection{$Z_{b}$ production in EIC and UPCs}

Photoproduction measurement is an important test of the structure of
exotic states. Since the energy span of EICs and UPCs facilities is large, it is
advantageous to find the exotic states in the bottom quark energy region.
Based on the previous $Z_{b}$ photoproduction results, with the help of
eSTARlight and STARlight programs, one obtains the cross-sections and event
numbers of $Z_{b}(10610)$ production in EICs and UPCs. As presented in Tab~\ref{cstable}, %
 the results based on the several accelerator devices are
calculated. The cross-section of $Z_{b}(10610)$ in UPCs
is larger than that in EICs since the photon flux of the nucleus is larger than the electron beam.
However, the event number of $Z_{b}(10610)$ in EICs
is larger than that in UPCs, especially FCC.

\begin{table}[h]
\caption{Cross sections and event numbers of $Z_{b}(10610)$ in $e$-$p$
scattering and $p\mbox{-}A$ UPCs. The integrated luminosities are the same
as Ref.\protect\cite{Klein:2019avl}. The luminosity $15\times10^{33}$ cm$%
^{-2}$ s$^{-1}$ was assumed for $10^7$ s of running was assumed for FCC
\protect\cite{Bordry:2018gri}.}
\label{table01}%
\scalebox{0.8}{
\begin{tabular}{|c|c|c|c|c|c|}
\hline\hline
& $e$-$p$ EIC-US & $e$-$p$ LHeC& $e$-$p$ FCC  & $p$-$Au$ RHIC & $p$-$Pb$ LHC \\ \hline
Beam energy, GeV & 18 (e) vs. 275 (p) & 60 (e) vs. $7\times 10^3$ (p) & 60 (e)  vs. $50\times 10^3$ (p) & 100 (p) vs. 100 (Au)
& $7 \times 10^3$ (p) vs. $2.778\times 10^3$ (Pb)  \\ \hline
Integrated luminosity & 10 $\mathrm{fb}^{-1}$ & 10 $\mathrm{fb}^{-1}$ & 150 $\mathrm{fb}^{-1}$   & 4.5 $\mathrm{pb}^{-1}$ & 2 $\mathrm{pb}^{-1}$ \\ \hline
$Z_b(10610) $ Cross sections &6.2 pb & 8.5 pb &9.8 pb & 2.0 nb & 30 nb \\ \hline
Expected statistics, $10^6$ events & 0.062 & 0.085 & 1.5& 0.0090  & 0.060  \\ \hline\hline
\end{tabular}}\label{cstable}
\end{table}

It can be seen from Tab.~\ref{cstable} that the differences between the total cross sections in EICs are small since
the cross-section of EICs is not very sensitive to the collide energies. However, the cross-section of $Z_{b}$ in EICs heavily depends on the value of the cross-sections of $Z_{b}$ photoproduction. It can be seen from Fig. \ref{tcs} that if the c.m. energy is too small, the photoproduced cross-sections of $Z_{b}$ will also be small, which indirectly leads to small cross-sections of $Z_{b}$ in EICs. For the situation of EicC \cite{Chen:2018wyz}, the present designation of colliding energies of EicC is 5 GeV vs 20 GeV, therefore, the c.m energy is about 16.7 GeV. Thus the cross-sections of $Z_b(10610)$ in EicC will be very small. Tab. \ref{tab3} shows the cross-section of $Z_{b}$ in EIcC and the number of the event when the beam energy is up to 10 GeV vs 100 GeV. The relevant results will provide theoretical references for future experimental measurements and upgrades.

\renewcommand\tabcolsep{0.26cm} \renewcommand{\arraystretch}{2}
\begin{table}[tbph]
\caption{Cross sections and event numbers of $Z_{b}(10610)$ in $e$-$p$ collision. Here, the integrated luminosity is taken as 10 $\mathrm{fb}^{-1}$, which is the current design value of EicC \cite{Chen:2018wyz}}
\label{tab3}%
\begin{tabular}{ccccc}
\hline\hline
Beam energy & 5 (e) vs. 20 (p) & 5 (e) vs. 30 (p) & 10 (e) vs. 50 (p) & 10 (e) vs. 100 (p) \\ \hline
$Z_b(10610) $ cross sections & 0.52 pb & 1.3 pb & 3.4 pb & 4.4 pb \\ \hline
Expected statistics, $10^3$ events & 5.2 & 13 & 34 & 44 \\
\hline\hline
\end{tabular}%
\end{table}

\begin{figure}[h]
\centering
\includegraphics[width=0.45\textwidth]{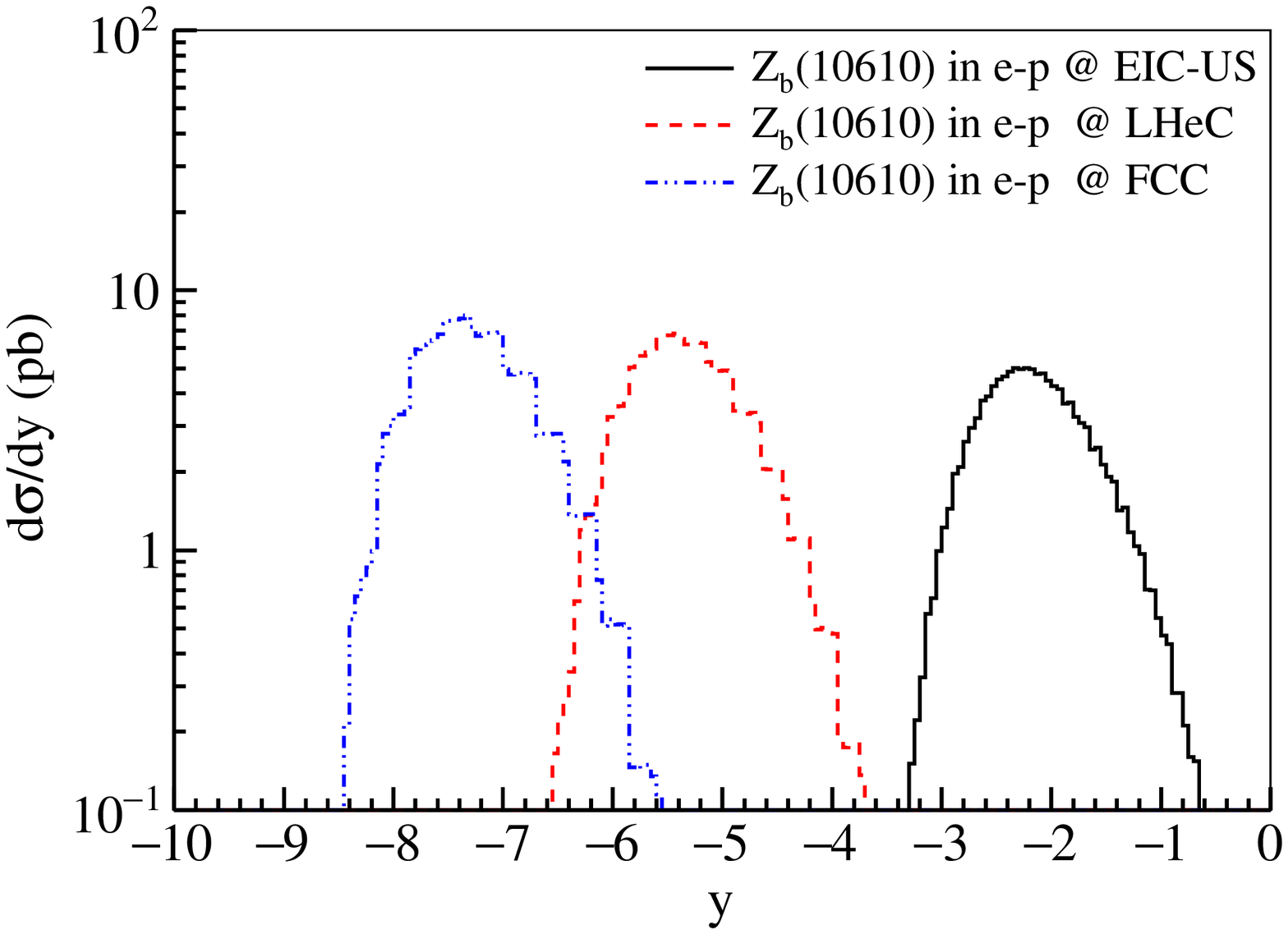} %
\includegraphics[width=0.45\textwidth]{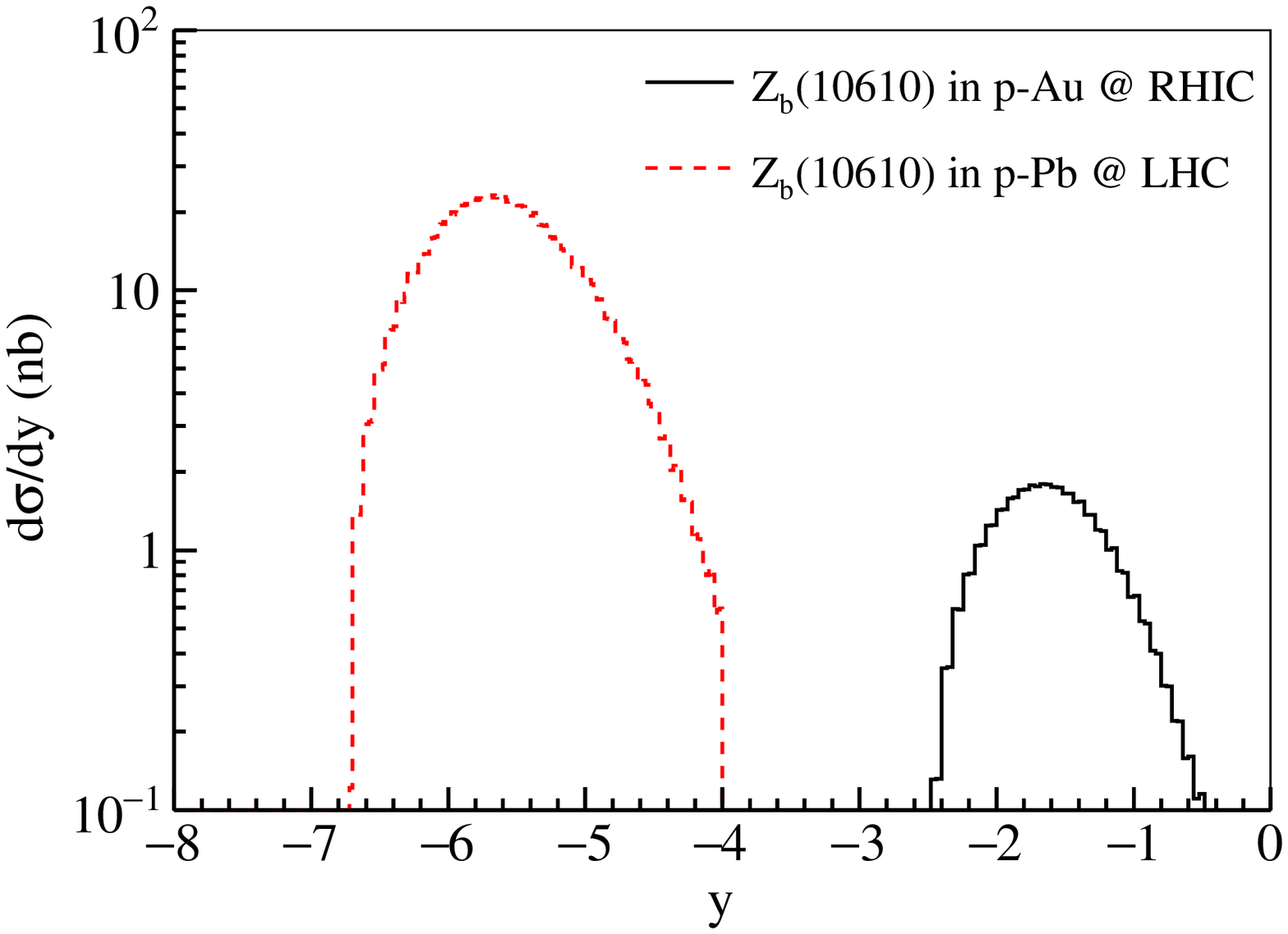}
\caption{(Color online) Rapidity distributions of $%
Z_{b}(10610)$ in e-p at EIC-US, LHeC and FCC with $0<Q^{2}<1~\mathrm{GeV}^{2}
$ and $p$-$Au$ and $p$-$Pb$ UPCs at RHIC and LHC. }
\label{rap}
\end{figure}

For the $Z_{b}(10610)$ production in EICs and UPCs processes, we also present the rapidity distributions of $Z_{b}(10610)$ in $e$-$p$ and $p$-$A$ processes, as shown in Fig.~\ref{rap}. These two distributions are relatively wide, which are related to the production mechanism of $Z_{b}$ through $t$-channel with pionic Regge trajectory exchange \cite{Wang:2015lwa}. This indicates that the distribution shape of the rapidity and transverse momentum reflects the shape of the cross-section in Fig.~\ref{tcs}. From the left graph in Fig.~\ref{rap}, it can be seen that photoproduction of EIC-US is near mid-rapidity and it is easy to identify in
the detector system. From the right graph of Fig.~\ref{rap}, we can also conclude that it is easy to observe $Z_b(10610)$
in $p$-$Au$ UPCs than $p$-$Pb$ UPCs since the $Z_b(10610)$ will be produced near mid-rapidity.

\begin{figure}[h]
\centering
\includegraphics[width=0.45\textwidth]{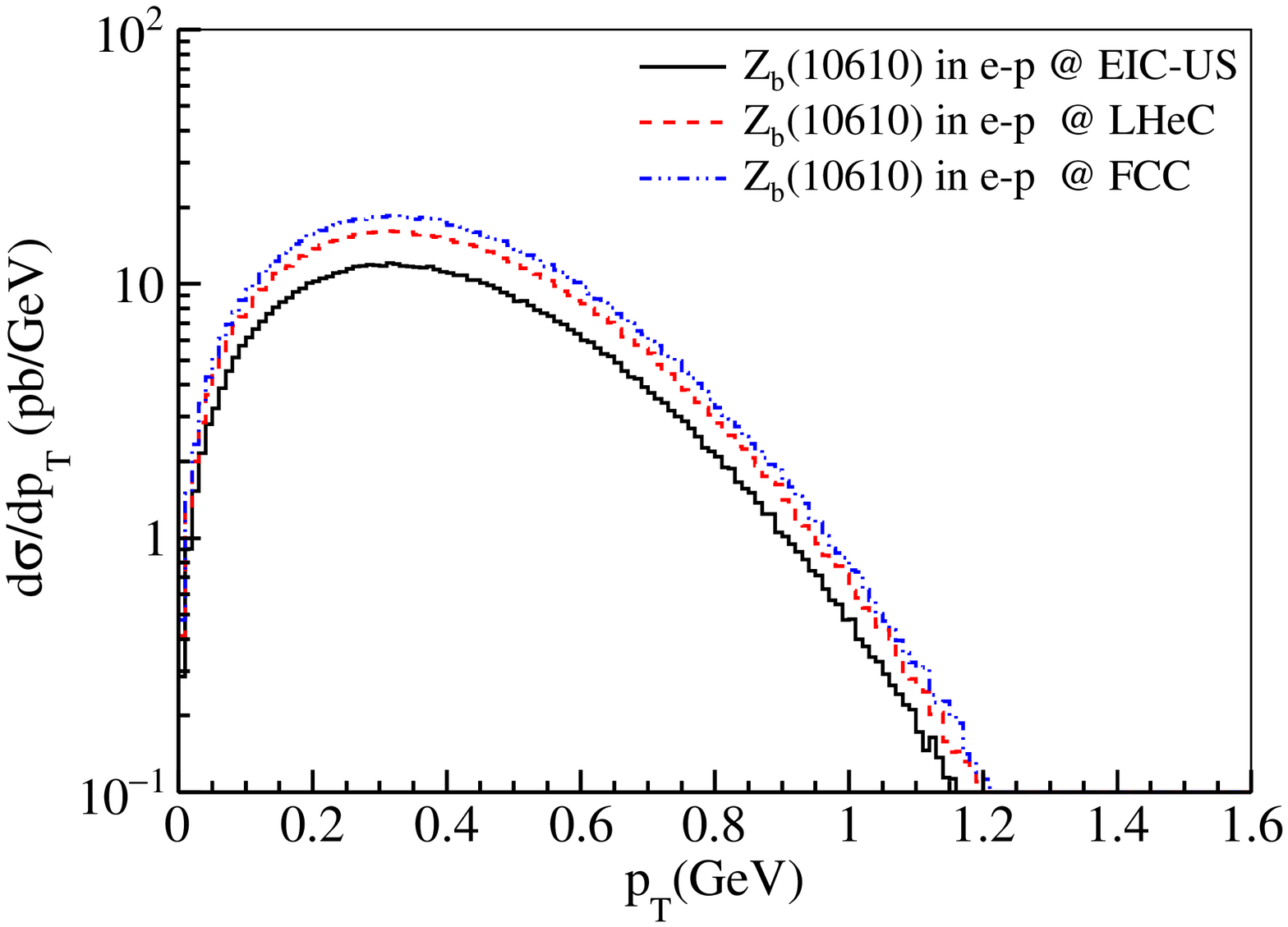} %
\includegraphics[width=0.45\textwidth]{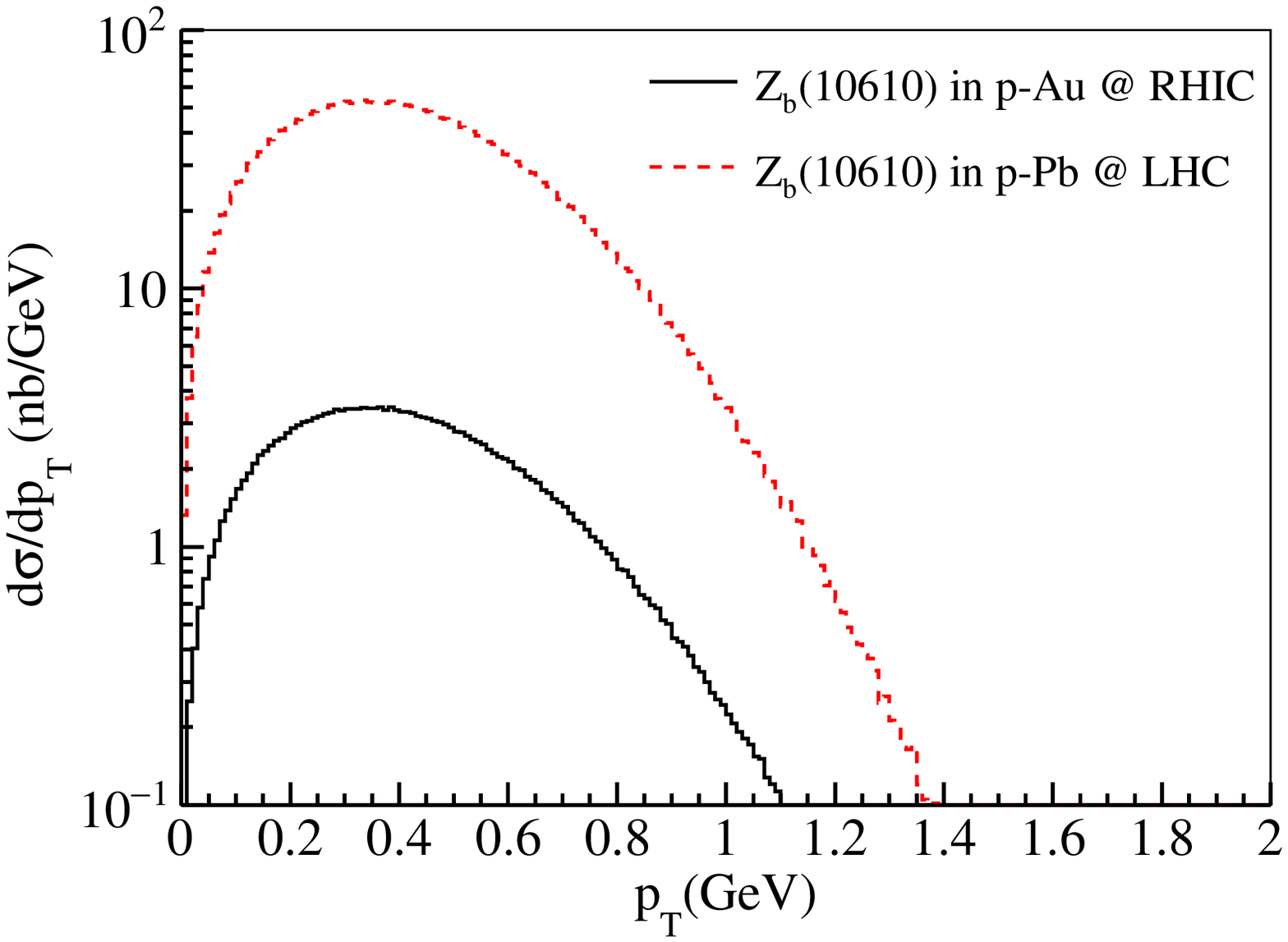}
\caption{(Color online) Transverse momentum distributions of $%
	Z_{b}(10610)$ in e-p at EIC-US, LHeC and FCC with $0<Q^{2}<1~\mathrm{GeV}^{2}
	$ and $p$-$Au$ and $p$-$Pb$ UPCs at RHIC and LHC.  }
\label{pt}
\end{figure}
In addition, the transverse momentum distributions in $e$-$p$ scattering and $p$-$A$ UPCs
are also shown in Fig.~\ref{pt}. These results can be
used as predictions for experiments. We notice that the three transverse momenta of
EICs are closed to each other since the total cross-sections are closed to each other. These
distributions can be employed to identifying the exotic states. In $p$-$A$ UPCs, the
difference between the two distributions is large because the total cross-section in $p$-$Pb$
is larger than the cross-section in $p$-$Au$. It can be founded that the largest value of
transverse momentum is about 0.2 - 0.4 GeV in both EICs and UPCs.

\begin{figure}[h]
\centering
\includegraphics[width=0.45\textwidth]{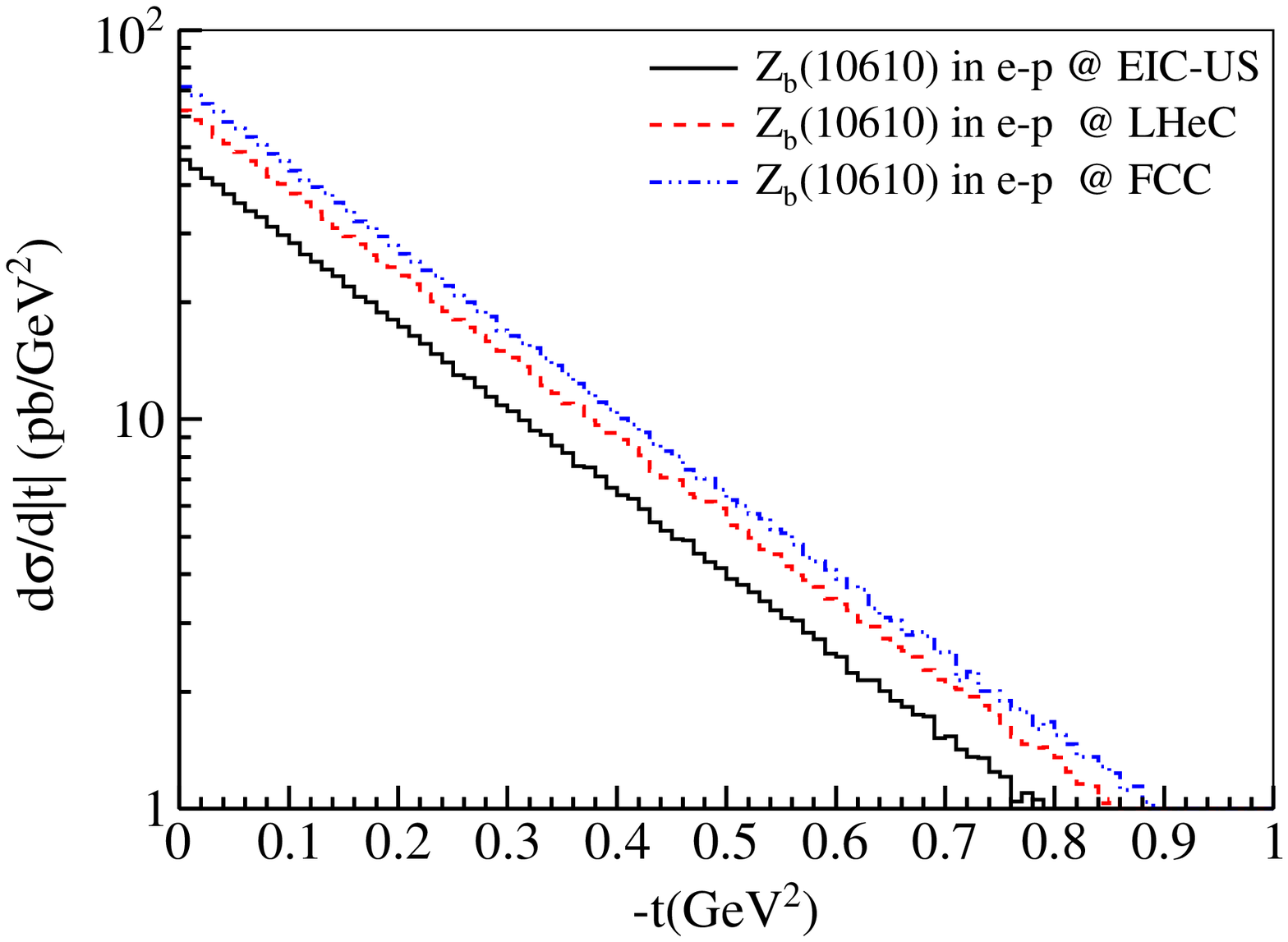} %
\includegraphics[width=0.45\textwidth]{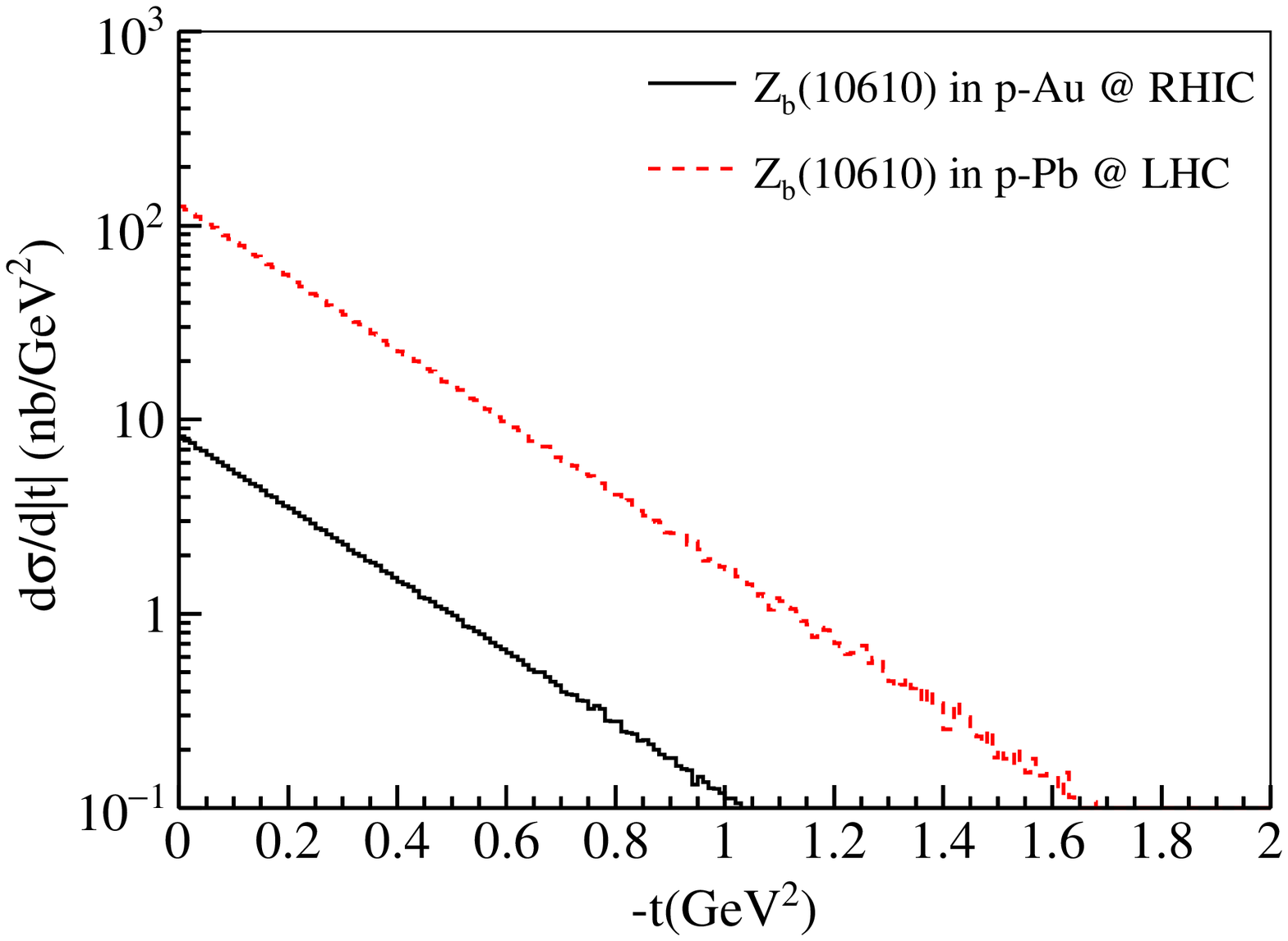}
\caption{(Color online) The $t$-distribution for the $Z_{b}(10610)$
production in $e$-$p$ and $p$-$A$ scatterings. }
\label{tslope}
\end{figure}

Moreover, We also give the $t$-distribution for the $Z_{b}(10610)$ production in $e$-$%
p $ and $p$-$A$ UPCs in Fig.~\ref{tslope}. The $t$-distributions are close to each other in three
EICs and are different in two UPCs. These conclusions are the same as the transverse momentum
distributions.
 Due to the adoption of the
Regge propagator, it can be seen from Fig.~\ref{tslope} that the shape of the curves of the differential
cross-sections of the $t$-distribution is relatively steep. This will be an important theoretical basis
for us to clarify the role and contribution of the Regge propagator through
EIC or UPCs experiments.

\section{Summary and discussions}

In this work, based on the effective field theory and the VMD mechanism, the
photoproduction of two bottomonium-like $Z_{b}(10610)$ and $%
Z_{b}(10650)$ states are investigated for the first time. The numerical results show that the total
cross-sections of the $\gamma p\rightarrow Z_{b}n$ reaction reach a maximum
at the center of mass energy $W\simeq 22$ GeV, which indicates that the
center of mass energy 22 GeV is the best energy window for searching for the
$Z_{b}$ states via $\gamma p$ scattering. Hence, an experimental study of
the bottomonium-like states $Z_{b}$ via the $\gamma p$ reaction is
suggested.

With the help of eSTARlight and STARlight packages, the cross-sections and
event numbers of $Z_{b}(10610)$ production in EICs and UPCs have been presented in this work.
As shown in Table~\ref{table01}, the EICs may collect more events due to
the larger luminosity. Moreover, we also simulated the rapidity and
transverse momentum distributions of $Z_{b}(10610)$ in $e$-$p$ scattering and $p$-$A$
UPCs processes. These results will provide an important basis for
estimating the production and studying properties of $Z_{b}$ in the RHIC, LHC, EIC-US, LHeC,
and FCC in the future.

Since the Reggeons are composed mostly of quarks, using Reggeons to prove
the distributions of sea quarks and anti-quarks in nuclei may be a
feasible method \cite{Klein:2019avl}. In this work, the
photoproduction of $Z_{b}$ is calculated by introducing the Regge exchange model, thus,
the numerical results will be beneficial for experimental studying
of the Reggeon model. Also, the cross-section of $t$ distribution
of $Z_{b}$ in different scattering processes are obtained, which will provide an
important theoretical reference for clarifying the role and contribution of
Reggeon.

\section{Acknowledgments}

This project is supported by the National Natural Science Foundation of
China (Grant Nos. 12065014, 11705076 and 11747160), and by the Strategic Priority
Research Program of Chinese Academy of Sciences, Grant No. XDB34030301. This
work is partly supported by HongLiu Support Funds for Excellent Youth
Talents of Lanzhou University of Technology. We also acknowledge the Natural Science Foundation of Fujian Province (Grant No. 2018J05007) and the Natural Science Foundation of Jimei University (Grant No. ZQ2017007).

\end{document}